\documentclass{article}
\usepackage{amstex,spconf,graphicx}

\newcommand{\w}{\omega}

\begin{document}

\title{Time-Evolution of the Power Spectrum\\
of the Black Hole X-ray Nova XTE J1550-564}
\name{}

\address{Lorenzo Galleani \\
Dipartimento di Elettronica,
Politecnico di Torino,\\
C.so Duca degli Abruzzi 24,
10129 Torino, ITALY\\
\\ Leon Cohen\\
City University of New York \\
695 Park Ave. New York, NY 10021 USA  \\ \\
Douglas Nelson,
\\U.S. Dept. of Defense, Fort Meade, MD, USA \\
\\ Jeffrey D. Scargle
\\
Space Science Division ,
NASA Ames Research Center \\
Moffett Field, CA 94035-1000}

\maketitle

\begin{abstract}
We have studied the time evolution
of the power spectrum of XTE J1550-564,
using X-ray luminosity time series data
obtained by the {\it Rossi X-Ray Timing Explorer} satellite.
A number of important practical fundamental issues
arise in the analysis of these data,
including dealing with time-tagged event data,
removal of noise from a highly non stationary signal,
and comparison of different time-frequency distributions.
We present two new methods to understand the time frequency variations,
and compare them to the {\it dynamic power spectrum}
of Homan et al. \cite{homan}
All of the approaches provide evidence
that the QPO frequency varies in a
systematic way during the time evolution of the signal.
\end{abstract}

\section{Introduction}

Recent years have seen dramatic developments in
our ability to study various parts of the universe
in wavelength regions inaccessible from the surface
of the Earth and using new methods of data acquisition,
primarily from NASA satellites.
Among the most fascinating are X ray data that are
suspected to be emanating from black holes.
The X-ray luminosities of such sources are
typically highly variable in a quasi-random
way, and most investigators take the
{\it power spectrum}\footnote{
Throughout, we loosely use the term spectrum,
but this function should
be carefully distinguished from the {\it energy spectrum},
that is the luminosity as a function of the energy or
wavelength of the radiation.}
of the time series data
as the diagnostic tool of choice.
In a number of cases the spectra
show peaks at well defined frequencies,
but with relatively broad profiles -- in
contrast to the very narrow peaks of
say pulsars, which are strictly periodic.
These broad power spectral features
indicate a roughly periodic behavior,
called {\it quasi periodic oscillations},
or QPO's.
A number of investigators have
reported time variations in these
power spectra.

Variations in the  spectra are of course of
immense interest because they indicate  possible changing
physical conditions and hence give basic information about
the presumed {\it accretion} process -- the
flow of matter from a disk of gas surrounding the
black hole into its deep gravitational potential well.
The basic presumption is that the variations
are connected with irregularities of the flow,
which is presumably turbulent.
Progress in this field is
crucially dependent on
proving that the power spectra are indeed
evolving, and studying the properties of
such variation.

This paper reports on new tools for studying
such spectral evolution, applies them
to data on a well known black hole candidate,
namely XTE J1550-564, discovered and
studied using NASA's RXTE Satellite.
The next section briefly describes the
standard tool used by astronomers, with
which our methods are to be compared.

\section{Dynamic Power Spectra}

Astronomers usually approach the question
of whether the harmonic and other frequency
content of time series are changing over time
by computing {\it dynamic power spectra}
(sometimes called {\it spectrograms})
from the time series data.
The point of view is that once a spectral feature
has been detected and studied by means of ordinary
power spectrum analysis \cite{VanDerKlis_sp},
then time-frequency analysis \cite{cohen}
is warranted, and the dynamic power spectrum
is the tool of choice.

We surveyed the Astronomical Journal and
found 15 papers using dynamic power spectra.
In all but two cases the method of computation
is essentially not stated, but we presume that
the authors have computed the ordinary power
spectrum in a sliding time-window.
This cavalier attitude -- as though
the method is so obvious and universal as to
be barely worthy of mention -- implies that
astronomers are largely unaware of the
shortcomings inherent in the dynamic power spectrum
\cite{cohen}.

We have developed two methods that we feel are
improvements over the usual tools used by astronomers,
but are nevertheless suited to the data modes in use in astronomy.

To exercise our methods we have obtained the raw data used in
generating the ``Dynamic Power Spectrum'' in Figure 19  of \cite{homan}.
This image shows a QPO initially located around 10 Hz,
apparently followed by a transition to lower frequencies, about 5-6 Hz.
The picture is particularly interesting because it suggests the existence
of a time-varying frequency with a non trivial behavior.

\begin{figure}[t]
\centerline{\includegraphics*[bb=300 200 530 400, scale=1]{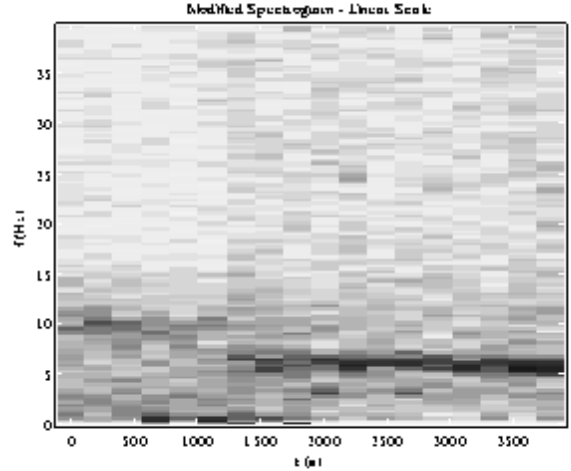}}
\caption{Reconstruction of the ``Dynamic Power Spectrum'' presented as Fig. 19
in \cite{homan}. The presence of a frequency component that makes
a transition between 10 Hz to 5-6 Hz is clearly visible.}
\label{DPS}
\end{figure}

We have been able to duplicate this figure.
Since the original data are photon arrival times, one has first to estimate
the underlying light intensity (rate of photon arrival).
This was apparently done by binning the data -- {\it i.e.} by
dividing the time axis in fixed intervals of constant length $T_s$
and counting the photons present in each.
We deduce that $T_s$ was $\frac{1}{80}$ seconds,
and that the instantaneous spectrum was estimated with
a Welch's periodogram.
Figure \ref{DPS} shows our reconstructed dynamic power spectrum,
which is essentially identical to that in \cite{homan}.

% Figure on the reconstructed Dynamic Power Spectrum

\section{A Time-Frequency Investigation of the X-ray Data}

Although binning is an intuitive and universally accepted density 
estimation procedure,
many other methods have more appealing properties \cite{Silverman},
especially considering that some additional signal processing has to be done
on the estimated density.  A famous approach is the kernel method
that consists in
averaging the number of events (photons) contained in
a relatively narrow window that moves continuously along the data.
The average is weighted with a normalized kernel that is more
or less concentrated about the center of the window.
Typical kernels are hence Gaussian, Hanning, triangular, etc.
The advantages in applying the kernel methods are several:

\begin{itemize}

\item
no dependence on phasing of the bins relative to the signal
(the dependence on bin size is replaced with that on the width of the kernel)

\item
the resulting density inherits properties of the window
({\it e.g.} it can be differentiated to the same order as the window itself)

\item
noise (and quantization) are suppressed

\end{itemize}

We therefore decided to apply the kernel method, using a normalized 
Hanning window.

Inspection of either density estimate indicates
the presence of slowly varying components,
on time scales $\ge 0.3$ second (corresponding to frequencies below 3 Hz).
To suppress this feature, with the objective to highlight the presence of 
the QPO,
we applied a high-pass filter to the data, using the forward-backward 
technique
with a simple elliptic filter.
This method eliminates the nonlinear phase distortion introduced by the filter
that can produce undesired effects in the time-frequency plane.

\begin{figure}[t]
\centerline{\includegraphics*[scale=.49]{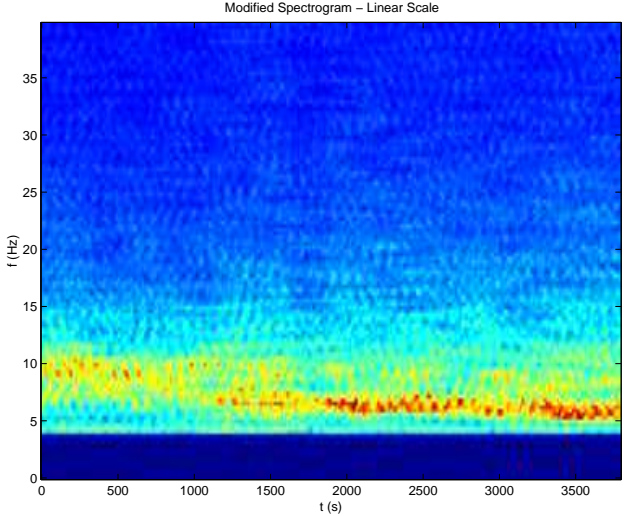}}
\caption{Time-frequency plot of the light intensity emanating from the black hole
obtained using the sliding estimator technique.}
\label{SlidingEstimator}
\end{figure}

At this point, because of the high noise level
(due almost entirely to Poisson counting fluctuations),
we estimate the instantaneous spectrum by applying
the so called ``sliding estimator'' \cite{rest} to the density.
This estimator is the natural extension to the spectrogram \cite{cohen},
and is computed by evaluating
Welch's periodogram of a windowed version of the signal.
The window is slid continuously along the signal,
and hence the dynamic power spectrum can be seen as a subcase of this method.
Fig. \ref{SlidingEstimator} represents the Time-Frequency plane obtained
with this technique.  (The more expressive color image can be found on the
CD rom of the proceedings or at the web page helinet.polito.it/sasgroup/ sliding.htm).
The joint use of kernel method, highpass filtering, and sliding estimator
confirms the presence of a time-varying spectrum in an outstanding way.

% Figure on the Sliding Estimator

In particular it is possible to notice the first concentration of the QPO 
around
10 Hz and then the transition to a central frequency around 6 Hz.
A more careful observation points out two interesting facts:

\begin{enumerate}

\item
the transition is a ``complex'' event, because it seems to involve a frequency
bifurcation around $t=800$ seconds;

\item
the QPO, especially after the transition for $t>1200$ s is well represented by
what is called a one component signal -- that is, a time-varying frequency 
with
slowly varying amplitude.  This can also be noticed in the first part of 
the QPO,
but with a smaller intensity.

\end{enumerate}

\noindent
It is also useful to observe how the noise is reduced, especially at 
higher frequencies,
by the application of the kernel method, and how the QPO is highlighted by
the highpass filtering.

\section{A Further Development in the Analysis of the X-ray Data}

\noindent
In the first part of this paper we have presented a nonparametric estimation of the
instantaneous spectrum of the intensity of light. This analysis has pointed out that
the QPO seems to be made by a time-varying frequency that also shows a transition
from the 10-11 Hz band to the 5-6 Hz band. To accomplish this analysis,
we adopted on purpose a nonparametric approach.
That is, we did  not make use of any model for the data,
because still we were not sure of the possible presence of oscillations in the signal.

Now that the result obtained by using the sliding estimator points out with a good
probability the existence of this oscillation, we intend to highlight it
by doing a further signal processing on the estimated time-frequency distribution.
In doing so we adopt a semi-parametric approach, using both additional
nonparametric methods and also a parametric model for the background noise
that will help us extracting the useful information.

For convenience we call $P(t,\w)$ the estimated instantaneous spectrum obtained
in the previous section, that we set as the starting point of the new analysis.
This analysis is based on three main steps: noise whitening, thresholding
and amplitude normalization of the QPO. A description of the steps follows,
in which we indicate with $P_k(t,\w)$ the time-frequency distribution obtained
at the $k$-th step applying the proposed signal processing.

\subsection{Noise Whitening}
One of the characteristics of the QPO that we have highlighted with the
sliding estimator technique, is that it is embedded in a high level non-white
noise, as can be confirmed by visual inspection of Fig. \ref{SlidingEstimator}.
This noise does not have a flat spectrum, and our aim here is to whiten it,
that is to force it to have a flat behavior.
%For noise region we mean the region outside the QPO, that is outside the
%5-11 Hz band.

Of course we know that in what we call noise, that is the spectral region outside
the QPO, there might be other important features of the black hole candidate
(for example other low energy QPOs), but as said before here we want
to focus our attention on the main QPO.
We accomplish the whitening in three steps.

\begin{itemize}

\item
We first estimate the power spectral density $|N(\w)|^2$ of the noise
by integrating with respect to time the time-frequency distribution $P(t,\w)$,
\begin{equation}
|N(\w)|^2=\int P(t,\w) dt
\end{equation}
This quantity is called frequency marginal in time-frequency analysis,
and can be seen as an approximation of the power spectrum of the signal.

\item
Then we fit the obtained noise density $|N(\w)|^2$ with a cubic polynomial
to obtain a fitted spectrum $|N_f(\w)|^2$,
paying attention to force the fitted curve to be grater or equal to a small
positive parameter $\delta$.
We point out that the noise density $|N(\w)|^2$ is estimated outside the QPO
region, whereas the fitted density $|N_f(\w)|^2$ is valid throughout the entire band.

\item
Finally we whiten the time-frequency distribution $P(t,\w)$ by dividing it
by the estimated noise
\begin{equation}
P_1(t,\w)=\frac{P(t,\w)}{|N_f(\w)|^2}
\end{equation}

\end{itemize}

\subsection{Thresholding}
Now that we have whitened the noise, and we expect it to show a flatter spectrum,
we threshold the obtained distribution $P_1(t,\w)$ to extract the QPO.
To do this we act in two steps

\begin{itemize}

\item
We estimate a threshold $\theta$ by inspecting the frequency marginal of $P_1(t,\w)$.

\item
We then subtract the threshold from the time-frequency distribution, setting to zero
any negative value the resulting difference may have
\begin{equation}
P_2(t,\w)=P_1(t,\w)-\theta, \quad P_2(t,\w) \ge 0
\end{equation}

\end{itemize}

\subsection{Amplitude Normalization of the QPO}
As a final step we want to compensate the instantaneous amplitude variations
of the QPO that can be noticed from Fig. \ref{SlidingEstimator}, and that are
probably responsible of the changing in time of the energy concentration
along the QPO instantaneous frequency.
To do this we normalize the instantaneous spectrum at each time by diving it
by its estimated energy at that time, that is
\begin{equation}
P_3(t,\w)=\frac{P_2(t,\w)}{\int P_2(t,\w) d\w}
\end{equation}
where the integral of $P_2$ over time is used as an approximation
to the energy of the instantaneous signal at that time (time marginal).

\begin{figure}[t]
\centerline{\includegraphics*[scale=.43]{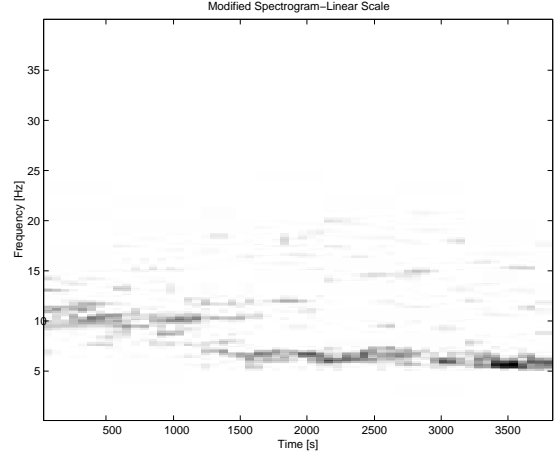}}
\caption{Extraction of the QPO with a more advanced and quasi-parametric signal
processing.}
\label{DougPicture}
\end{figure}

In Fig. \ref{DougPicture} we plot the final instantaneous spectrum $P_3(t,\w)$,
where it is possible to notice the effects of the above procedure.
The analysis done so far better highlights what seems to be
the time-varying behavior of the QPO central frequency.
Our analysis to date seems to elucidate QPO variations better than
conventional dynamic power spectra, but further studies are needed, 
mainly of the algorithm's reliability and immunity to artifacts.

\vskip .2in
%\small
\noindent{\em Acknowledgment:}
Work supported by the NSA HBCU/MI program
and the NASA Applied Information Systems Research Program.

\end{document}